\documentclass[prd,twocolumn,a4paper,altaffilletter,showkeys]{revtex4}

\usepackage{graphicx}
\usepackage{dcolumn}
\usepackage{bm}
\usepackage{amsmath,amssymb}
\usepackage{euscript}

\newcommand{\be}{\begin{equation}}
\newcommand{\ee}{\end{equation}}
\newcommand{\bea}{\begin{eqnarray}}
\newcommand{\eea}{\end{eqnarray}}

\def\vh{\varphi}

\newcommand{\eu}{\EuScript}

\begin{document}

\title{\bf  Hamiltonian Cosmological Dynamics of  General Relativity}

\author{B.M. Barbashov, V.N. Pervushin, V.A. Zinchuk, and  A.G. Zorin
}
\affiliation{Bogoliubov Laboratory of Theoretical Physics,\\
Joint Institute for Nuclear Research, 141980 Dubna, Russia}

\date{\today}

\begin{abstract}
 {\small { The Hamiltonian approach to General Relativity  is
 developed similarly to
 the Wheeler-DeWitt Hamiltonian cosmology,
 where the cosmological scale factor is treated as a time-like
 dynamic variable and its canonical momentum is considered as
 an evolution generator in the field space of events
 with the postulate about a physical vacuum as a state with
 the minimal eigenvalue of this generator.

  The cosmological scale factor is extracted from
  the Hamiltonian General Relativity without
  double counting of the spatial metric
determinant in  contrast to the standard cosmological perturbation
theory.
   The Friedmann-like equations in the exact
theory are  derived. A new version of cosmological perturbation
theory
  keeps the form
 of the  Newton interactions in an early Universe.
  We show how the considered  Hamiltonian approach to GR
  can solve the topical problems of modern cosmology and quantum
  theory of gravitation.
 }}
\end{abstract}


\maketitle

\newpage

\section*{Introduction}

 The status of the cosmological scale factor  in the
modern theory
 is ambiguous. The standard  Hamiltonian approach to General
 Relativity (GR)  \cite{dir, ADM, berg, shw, fadpop} ignores the scale factor
  by the
 choice of the corresponding class of functions where the gauge of
 minimal surface \cite{dir} with the unit scale factor   is possible. The
 Wheeler - DeWitt Hamiltonian cosmology \cite{M,WDW,Ryan1,Ryan2}
 considers the scale factor as an independent dynamic variable.
 In the standard cosmological perturbation theory  \cite{lif,bard,kodam}
 the scale factor is treated rather as the external homogeneous field than
an independent dynamic variable.
 From this point of view all these three branches of GR appear as three
 different theories.

  The questions arise: What is the version of the Hamiltonian GR
  defined on the class of functions that includes the cosmological  scale factor?
  What is the
 version of the  cosmological perturbation theory,
 where the  scale factor plays
 the role of an independent dynamic variable?

In the present paper, to answer  these questions, we use a deep
analogue between GR and Special Relativity (SR) \cite{WDW} with
the field space of events as the generalization of the Minkowski
space of SR and the cosmological scale factor playing the role of
a time-like variable in this field space.

 This time-like  variable and the corresponding
 Hamiltonian dynamics  can be revealed in a specific frame of
 reference defined in \cite{dir, ADM, berg, shw, fadpop}.
 Fixation of this specific frame is associated with
 a set of instruments for measuring
 fields and keeps  the number of variables in
 contrast to fixation of gauge constraints decreasing the number of dynamic
 variables.

 To separate the  gauge transformations
 from the frame ones, it is natural to use the representation
 of the geometric interval in terms of the
  Cartan forms introduced in GR by Fock \cite{fock29}.
  The Cartan forms are gauge invariant and
 relativistic covariant \cite{og}.

 The content of the paper
 is the following.

 In Section \ref{hr},
    the Dirac Hamiltonian approach to GR is considered in terms of
 the Cartan forms in order to separate the frame transformations
 from the gauge ones.

In Section \ref{cs}, arguments are listed in favor of the
 solution of the topical problems of the Hamiltonian
approach to GR by the choice of an evolution parameter  in the
field space of events as the cosmological scale factor.

 In Section \ref{chr}, the Hamiltonian approach to GR with
 the cosmological dynamics is developed
 as a new version of cosmological perturbation theory.

\section{\label{hr}Dirac Hamiltonian approach to General Relativity}

\subsection{The action, metric, and symmetry}

 In order to state problems, we  consider the Dirac
 approach \cite{dir} to the Einstein-Hilbert action
 \be\label{gr}
 S_{GR}[\varphi_0|e]=-\int d^4x \sqrt{-g}\frac{\varphi^2_0}{6}R(g),\ee
where
 $$
 \varphi^2_0=\frac{3}{8\pi}M^2_{\rm Planck},~~~ \hbar=c=1,
 $$
 given in the space with the
 interval
 \be \label{ds}
 ds^2=\eta^{\alpha\beta}\omega_{(\alpha)}\omega_{(\beta)}
 \equiv g_{\mu\nu}dx^\mu dx^\nu,\ee
 \be
 \eta^{ab}=diag(1 -1 -1 -1),
 \ee
where
 \be\label{raz} \omega_{(\alpha)}=e_{(\alpha)\mu}dx^{\mu},
 \ee
  are the linear Cartan forms  \cite{fock29,fadpop}. These forms
   allow us to include fermions and other fields $f$,
  if  Standard Model will be added
  \be\label{t1}
 S[\varphi_0|e,f]=S_{\rm GR}[\varphi_0|e]+S_{\rm SM}[\varphi_0|e,f],
 \ee
    and separate the gauge transformations from the frame transformations
  \cite{og}.
   In particular, the Cartan forms $\omega_{(\alpha)}$ are
 invariant under the general coordinate transformations
 \be\label{gct}
 x^{\mu}~~~~~\longrightarrow~~~~~\tilde{x}^{\mu}=\tilde{x}^{\mu}(x^0,x^1,x^2,x^3)
 \ee
 treated as  gauge ones (that are accompanied  by the
 constraints).
 The Cartan forms $\omega_{(\alpha)}$
 covariant under the Lorentz  transformations of type
 \be\label{lt}\left\{\begin{aligned}
 \bar{\omega}_{(0)}&=\frac{\omega_{(0)}-V\omega_{(1)}}{\sqrt{1-V^2}},\\
 \bar{\omega}_{(1)}&=\frac{\omega_{(1)}-V\omega_{(0)}}{\sqrt{1-V^2}},\\
 \bar{\omega}_{(2)}&=\omega_{(2)},\\
 \bar{\omega}_{(3)}&=\omega_{(3)}\\\end{aligned}\right.
 \ee
 treated as transformations of frames of references. Recall that
 the latter (i.e., frame transformations) are associated
 with conservation numbers and keep
 number of variables, whereas the
 first (i.e., gauge ones) lead to constraints \cite{dir} that
 decrease the number of variables.

\subsection{Frame of reference}

A choice of a Lorentz frame in GR   means the fixation
 of the  Lorentz indices ${(\alpha)}$ in the Cartan forms (\ref{raz})
 and their classification into the time-like $\omega_{(0)}$ and space-like ones
 $\omega_{(a)}$.

 The Hamiltonian dynamics is  formulated in the specific Lorentz frame
 using the Dirac -- ADM $3+1$ foliation of  the Cartan forms
 \cite{dir, ADM,vlad}
 \bea \label{adm}
 &\omega_{(0)}=Ndx^0,\\
 &\omega_{(a)}= e_{(a)i}(dx^i+N^i dx^0);\label{psi}
 \eea
 here triads $e_{(a)i}$ form the spatial metrics
 $$g^{(3)}_{ij}=e_{(a)i}e_{(a)j};
 ~~~ g^{(3)ij}{=}e^j_{(a)}e^i_{(a)}.
 $$
  Following  Dirac \cite{dir} one can factorize
  the determinant of the spatial
  metrics eliminating factor $\psi^2$ from triads
 \bea \label{d1}
 & e_{(a)i}=\psi^2 {\bf e}_{(a)i},~~~~~~~~ \det |{\bf e}|=1,\\
 &N=N_{\rm d}\psi^6.~~~~~~~~~~~~~~~~~~~~~~~~~~~~~
 \label{d2}
 \eea

One can use the symmetric parametrization of the Cartan forms
 $
{\bf w}_{(a)}={\bf{e}}_{(a)i}dx^i
 $ as a
 nonlinear realization of equiaffine
symmetry~\cite{og}.

 The Dirac-ADM parametrization characterizes
 a fami\\ly of hypersurfaces $x^0=\rm{const.}$ with the unit normal
 vector
 $\nu^{\alpha}=(1/N,-N^k/N)$ to a hypersurface.
 The second (external) form
 \begin{multline}\label{sc}\pi_{(a)i}=\frac{1}{N_d}
 \left[(\partial_0-N^l\partial_l)e_{(a)i}- e_{(a)l}\partial_i
 N^l \right]=\\=\psi^2\left[{\bf d}_{(a)i}+2{\bf
 e}_{(a)i}\pi_{\psi}\right],
 \end{multline}
 where
 \begin{multline}\label{proizvod}{\bf d}_{(a)i}=
 \frac{1}{N_d}\left[(\partial_0-N^l\partial_l){\bf e}_{(a)i}+\right.
 \\\left.+ \frac13 {\bf
 e}_{(a)i}\partial_lN^l-{\bf e}_{(a)l}\partial_iN^l\right] \end{multline}
 and
 \be\label{proizvod1}
 \pi_{\psi}=\frac{1}{N_d}\left[
 (\partial_0-N^l\partial_l)\ln{\psi}-\frac16\partial_lN^l\right],
 \ee
 shows us how this hypersurface is embedded into the
 four-dimensional
 space-time.

 The internal 3-dimensional scalar  curvature ${}^{(3)}R(e)$
  after   transformation (\ref{d1}) takes the form
 \be
 {}^{(3)}R(e)=\frac1{\psi^4}{}^{(3)}R({\bf e})+
 \frac{8}{\psi^5}\triangle\psi,
 \ee where ${}^{(3)}R({\bf{e}})$ is the curvature in
 terms of triads: ${\bf e}_{(a)i}$.

\subsection{The action in terms of the Dirac variables}

 In such a way we can decompose the action (\ref{gr}) into three
 terms: kinetic ${\bf K}$, potential ${\bf P}$, and surface
 ${\bf S}$
 \begin{multline}\label{l} S_{GR}[\vh_0|e]=\\
 =\int dx^0d^3 x \left({\bf K[\vh_0|e]-P[\vh_0|e]
 +S[\vh_0|e]}\right),
 \end{multline}
 where
 \bea
 {\bf K}[\vh_0|e]&=&N_d\varphi_0^2
 \left[\frac{1}{24}\left({\bf d}_{(a)j}{\bf e}^j_{(b)}+
 {\bf d}_{(b)i}{\bf e}^i_{(a)}\right)\times
 \right.\nonumber\\
  &&\left.\times\left({\bf d}_{(a)j}{\bf e}^j_{(b)}+
 {\bf d}_{(b)i}{\bf e}^i_{(a)}\right)
 -4\pi^2_{\psi}{\vphantom{\frac12}}\right],
 \label{k}\\
 {\bf
 P}[\vh_0|e]&=&\frac{N_d\varphi_0^2\psi^{12}}{6}~~{}^{(3)}R(e),
 \label{p}\\
 {\bf S}[\vh_0|e]&=&2\varphi_0^2\left[\partial_0\pi_{\psi}-
 \partial_l(N^l\pi_{\psi})\right]-\nonumber\\
 &&~~~~~~~~~~-
 \frac{\varphi_0^2}3 \partial_j[\psi^2\partial^j (\psi^6
 N_d)]\label{s};
 \eea
here we used the definitions (\ref{proizvod}) and
(\ref{proizvod1}).

 The equations obtained from the action (\ref{l})  have
 ambiguous solutions for metric components that depend on the
 initial data and gauge. To remove  gauge
 ambiguity, one needs to fix coordinates by  gauge constraints.
 Following  Dirac \cite{dir} we  fix physical coordinates
 by the constraint of transversality:
 \be\label{tr}
 \partial_i {\bf e}^{i}_{(a)}=0
 \ee
 and the minimal surface \cite{dir}:
 \be
 \pi_{\psi}=0\label{gauge}, \ee
 where $\pi_{\psi}$ is given by (\ref{proizvod1}).

\subsection{The Dirac Hamiltonian and gauges}

 The Dirac-ADM parametrization of the metrics  \cite{dir,ADM} leads to
 the GR action in the Hamiltonian approach
 \begin{multline}\label{ha} S_{\rm
 GR}[\vh_0|e]=\\
 =\int dx^0\left\{\int d^3x
 \left[p^k_{(a)}\partial_0{\bf e}_{(a)k}-p_\psi\partial_0\psi\right]-H_d\right\},
 \end{multline}
 where
 \begin{multline}\label{ha1}
 H_d= \int d^3x\left[ N_d{\cal H}_d(\vh_0| e)
 -N_{(a)} {\cal P}_{(a)}-\right.\\
 \left.-C_0p_\psi- C_{(a)}\partial_k{\bf e}^k_{(a)}\right]
 \end{multline}
 is the Dirac Hamiltonian; $N_d, N_{(a)}, C_0, C_{(a)}$ are
 Lagrangian multipliers,
 \be\label{hd}
 {\cal H}_d(\vh_0| e)= \frac{1}{\vh_0^2}
 \left[6p_{(ab)}p_{(ab)}-\frac{p^2_\psi}{16}\right]+
 \frac{\vh_0^2\psi^{12}}{6} {}^{(3)}R(e),
 \ee
 \be {\cal P}_{(b)}=
 \partial_jp^j_{(b)}+
 \left[p^j_{(a)}{\bf F}_{(a)kj}-T^0_{k}\right]{\bf e}^k_{(b)}
 \ee
 are the local Hamiltonian density and
  the local momentum, where we distinguished the energy momentum tensor
  depending on the trace of the second form
 \be \label{t0k}
 T^0_k= p_{\psi}\partial_k \psi-\frac{1}{6}\partial_k
 (p_{\psi}\psi) ~
 \ee
 and used the notation
 \be
p_{\psi}=\frac{1}{N_d}\frac{\partial {\bf K}[\vh_0|e]}{\partial
\pi_{\psi}},~~~~ p^i_{(a)}=\frac{\partial {\bf
K}[\vh_0|e]}{\partial(\partial_0{\bf e}_{(a)i})}
 \ee
 \be \label{nots}
 p_{(ab)}=\frac{1}{2}\left[p^i_{(a)}{\bf e}_{(b)i}+p^i_{(b)}{\bf
 e}_{(a)i}\right],\ee \be
{\bf F}_{(b)i j}=\partial_i{\bf e}_{(a)j}-\partial_j{\bf
e}_{(a)i}.
 \ee
 Variation of the action (\ref{ha}) under the Lagrange multipliers
 $N_d,~N_{(a)}$ leads to the first class
 constraints
 \be\label{f1}
 {\cal H}_d=0,~~~~~~~~~
 {\cal P}_{(a)}=0,
 \ee
and variation with respect to $C_0,~C_{(a)}$ leads to the second
class constraints
 \be\label{f2}
 p_\psi=0,~~~~~~~~~
\partial_k{\bf e}^k_{(a)}=0.
 \ee
 The conservation of the minimal surface
 \be\label{d0}
 \partial_0p_\psi= \left\{H_d,p_\psi\right\}=0
 \ee
 means the equation of motion of the spatial
 metric determinant  which we denote
  formally as the variation of the action with respect to $\log\psi$
\begin{multline}\label{d01}
 e^i_{(a)}\frac{\delta S}{\delta e_{(a)i}}=\frac{1}{2}\frac{\delta S}{\delta\log\psi}=
 -4{\bf P}[\vh_0|e]+2{\bf S}[\vh_0|e]=0,
 \end{multline}
  where ~ ${\bf P}[\vh_0|e],  ~ {\bf S}[\vh_0|e]$  are defined by Eqs.
  (\ref{p}) and (\ref{s}) where $\pi_\psi=0$.
 Note that Eq. (\ref{d01}) defines the differential
  operator $\hat{\bf A}[\vh_0|e]$:
  \be\hat{\bf
  A}[\vh_0|e]N_d\equiv-4{\bf P}[\vh_0|e]+2{\bf S}[\vh_0|e].
 \ee
 The Poisson brackets of the constraints take the forms \bea\label{d02}
 \left\{p_\psi,\int d^3y {\cal H}_d F\right\}&=&2\hat{\bf A}[\vh_0|e]F, \\
  \left\{\partial_k{\bf e}^k_{(a)},\int d^3y
 {\cal P}_{(b)}F_{(b)}\right\}&=&\hat{\bf B}_{(a)(b)}F_{(b)},\label{d03}
 \eea
 where
  \be
 \hat{\bf B}_{(a)(b)}F_{(b)}=\partial_{(b)}\partial_{(a)}F_{(b)};
 \ee
 here we used  the notation
 $\partial_{(a)}f={\bf e}_{(a)i}\partial_i f$.

\subsection{Problems of the Dirac Hamiltonian approach}

 Using the Poisson brackets (\ref{d02}), (\ref{d03}) one can
 formally write the Faddeev -- Popov (FP) functional integral \cite{fadpop}
 over
 the set of variables $f=(\psi,{\bf e}_{(a)i})$, their canonical
 momenta $p_f=(p_\psi,p^i_{(a)})$, and the Lagrange multipliers
 $C=(C_0,C_{(a)})$, $N=(N_d,N_{(a)})$
 \begin{multline} \label{z}
 Z_{FP}=\int \prod_{x}\left[ dC dN
 \left(\prod^{}_{f}\frac{dp_fdf}{2\pi}
 \right)\right]\\ \times (\det\hat{\bf A})(\det
 \hat{\bf B})e^{i(S+s.t.)}
 \end{multline}
 with the surface
 term (s.t.) treated as an energy.
 Eq. (\ref{d01}) $\hat{\bf A}N_d=0$ shows us that $\det \hat{\bf
 A}=0$, and
 it is a problem of the Gribov
 copies of the minimal surface gauge.

Faddeev and Popov proved \cite{fadpop, popkon} that
 the integral  (\ref{z}) is not equivalent to the one in relativistic
 invariant harmonic gauge $\partial_{\mu}(\sqrt{-g}g^{\mu\nu})=0$
 which does not depend on a frame of reference. This means that
 the  functional integral (\ref{z}) with the minimal surface
 depends on the frame of reference.
 Such the dependence does not contradict to relativistic
 covariance.
 (Recall that according to the theory of unitary irreducible representations
  of the  Poincare groups (see \cite{shwb})
 the relativistic invariance means
 the invariance of a complete set of frames of reference with
 respect to the Lorentz transformations.
 Therefore   only the complete set of
 functional integrals with minimal surfaces repeated in each frame
 of reference is relativistic invariant.)

 It is well known \cite{fadpop} that the Dirac formulation of
 quantum theory faced also  the problems of non-localizable energy, arrow
 of time, ultraviolet divergences, singularity, and initial data.
 Moreover, the minimal surface constraint contradicts  the
 observational data of the Hubble expansion rate $H$ because
 $p_{\psi}\sim H \neq 0$. There is an opinion that  including
 the cosmological scale factor as an evolution parameter allows us
 to solve these problems \cite{pp, bpp}. The
 necessity of the similar
 evolution parameter in Special Relativity  and cosmology
  follows from
 the invariance of GR under reparametrizations of the coordinate
 time (\ref{gct}).

\section{\label{cs} The Wheeler - DeWitt SR/GR correspondence}

\subsection{Time as a variable in Special Relativity }

  The dynamics of a relativistic particle is determined by the  action
 \be \label{sr}
 S_{\rm SR}=-\int d\tau [-p_{\mu}\partial_\tau x^\mu
  +\frac{e_{(1)}}{2m}(p_{\mu} p^{\mu}-m^2)]
 \ee
  given in the space of events $[x^0|x^i]$ and
 the proper time interval $ds=e_{(1)}d\tau$.
 Both the action and  interval are invariant under
 reparametrization of the coordinate time denoted here as
 $\tau\rightarrow\tilde{\tau}=\tilde{\tau}(\tau)$.
 These reparametrizations are treated
 as gauge transformations that lead to the mass-shell constraint
 $p_{\mu} p^{\mu}-m^2=0$.
 That means that
 there are no instruments for measurement of the coordinate time
 $\tau$.

 It is known \cite{poi,ein} that there are two measurable
 times in SR: the time as a variable $x^0(\tau)$, and the time as an
 interval $ds=e_{(1)}d\tau$. The time as the variable is revealed
 when the mass-shell constraint $p_{\mu} p^{\mu}-m^2=0$
 is solved in the specific frame with respect to
 $p_0$ ($p_0=\pm\sqrt{p_i^2+m^2}$) treated as an energy in
 the space of events. To remove
 the negative  values of the energy, one  postulates
  the existence of a vacuum as a state with minimal energy.
 This postulate restricts the region of the motion of a particle
 in the space of events, so that for the positive energy
 $p_{(+)0}=+\sqrt{p_i^2+m^2}$ a particle goes forward $x^0>x_I^0$,
 and for the negative energy
 $p_{(-)0}=-\sqrt{p_i^2+m^2}$ a particle goes backward $x^0<x_I^0$,
 where $x_I^0$ is treated as the initial data of the time-like
 variable. (In quantum theory the initial point $x_I^0$ is treated
 as a point of creation of a particle with positive energy $p_0\geq 0$,
  or
as a point of annihilation  of a particle with positive energy
when the energy of events is decreased $p_0\leq 0$.)

 The motion of a particle with positive energy in  the space of events
 can be described by
 the reduced action
 \be\label{srr}
 S_{\rm SR}{}_{{(\rm energy ~constraint)}}=
 -\int dx^0 [p_{i}\partial_0 x_i-\sqrt{p_i^2+m^2}]
 \ee
 defined as the action (\ref{sr}) on the
 constraint $p_{(+)0}=+\sqrt{p_i^2+m^2}$
 depending on a frame of reference.
 We can see that in the reduced action one of the dynamic variables
 ($x^0$) in the space of events $[x^0|x^i]$ plays the role of the
 physical evolution parameter, while  its momentum $p_0$ is the
 corresponding generator of evolution.

 However,  the reduced action (\ref{srr}) loses
 a geometric interval $ds=e_{(1)}d\tau$ with the coordinate time
 $\tau$, whereas action (\ref{sr}) contains the relation between
 the {\it dynamic evolution parameter} $x^0$ and the {\it geometric
 interval}  $s$.
 This relation can be obtained by varying  the action (\ref{sr})
 with respect to the momentum\,$p_0$
 \be\label{2so} \frac{\delta S_{\rm SR}}{\delta
 p_0}=0~~~\Longrightarrow~~~
 ds=e_{(1)}d\tau=\pm\frac{dx^0}{\sqrt{p^2+m^2}}m. \ee
 Thus, the complete
 description of a relativistic particle can be given by
 two equivalent unconstrained systems: the dynamic (\ref{srr}) and
 the geometric one 
 \cite{18}.
 As it was proposed by
 Wheeler and DeWitt \cite{WDW}, it is just the way to solve the similar
 problems in GR.

\subsection{Reparametrization-invariance in GR}

A gauge group of the Hamiltonian approach in the specific frame of
reference is considered as a
 group of  diffeomorphisms ~\cite{vlad} of the Dirac-ADM parametrization
 of the metric~(\ref{adm}), (\ref{psi})
 \be \label{gt}
 x^0 \rightarrow \tilde x^0=\tilde x^0(x^0);~~~~~
 x_{i} \rightarrow  \tilde x_{i}=\tilde x_{i}(x^0,x_{1},x_{2},x_{3})~,
 \ee
 \be \label{kine}
 \tilde N = N \frac{dx^0}{d\tilde x^0};~~~~\tilde N^k=N^i
 \frac{\partial \tilde x^k }{\partial x_i}\frac{dx^0}{d\tilde x^0} -
 \frac{\partial \tilde x^k }{\partial x_i}
 \frac{\partial x^i}{\partial \tilde x^0}~.
 \ee
 These transformations
 conserve the family of hypersurfaces $x^0=\rm{const.}$, and they
 are called a kinemetric subgroup~\cite{vlad, ps1}
 of the group of general coordinate
 transformations~(\ref{gct}).
 The group of kinemetric transformations contains
 reparametrizations of the {\it coordinate time}~(\ref{gt}).
 This means that
 there are no physical instruments that can measure
 this {\it coordinate time} $x^0$. That requires  introducing the
 evolution parameter as one of dynamical variables,
 as we have seen in SR. This time-like dynamic variable is
 identified with the scale factor in the Wheller -- De Witt (WDW)
  Hamiltonian cosmology \cite{WDW}.

\subsection{The  WDW Hamiltonian cosmology }

 In the Hamiltonian cosmology one uses
  the Wheeler--Dewitt SR/GR correspondence
 between coordinate times $[\tau/x^0]$, dynamic variables $[x^0|x^i]/[\psi|F]$,
 and gauge symmetries $[(\tau\to \widetilde{\tau})/
  (x^0\to \widetilde{x}^0)]$ \cite{WDW}.
 Wheeler and DeWitt  proposed to consider the spatial
 metric determinant $\psi$ as the {\it dynamic evolution parameter}.
 In the homogeneous approximation
  this time-like variable $\psi^2(x^0,x^i)$ becomes the
 cosmological scale factor $a(x^0)$.

  The WDW cosmology with the {\it dynamic evolution parameter}
 is defined as a homogeneous approximation of the
theory (\ref{t1}) in the space-time with the interval
 \be\label{cfm}
ds=a^2(x^0)[N_0^2(x^0)(dx^0)^2-(dx^i)^2],
 \ee
 where  the Hamiltonian  $H_d$ in the action
(\ref{ha}) is replaced by its expectation value $V_0\rho$.
  In this case, the action (\ref{ha}) reduces to
  a constrained mechanical system of the type of SR \cite{WDW,M,Ryan1,Ryan2}
 \be\label{ham} S_{c}=\int
 dx^0\left[-P_\vh\partial_0
 \vh+N_0\left(\frac{P^2_\vh}{4V_0}-\rho_0(\vh)V_0\right)\right],
 \ee
 where $\vh(x^0)=\vh_0a(x^0)$, $\rho_0(\vh)$ does not contain
 internal dynamic variables except for $\vh$,
 and the lapse function $N_0$
 plays the role of the Lagrange
 multiplier.
 The equation of $N_0$
 \be\label{conn}
 \frac{P^2_\vh}{4V_0}-\rho_0(\vh)V_0=0 \ee
 is the energy constraint with the  solution
  \be \label{10}
 P_\vh=\pm E, \ee where \be \label{110}
 E=2V_0\sqrt{\rho_0(\vh)}.\ee

The theory  (\ref{ham}) was considered in a similar way as the
theory of a relativistic particle in the  space of events
\cite{WDW,M} with the cosmological scale factor $\vh$  defined  as
a time of events, $P_\vh$ considered as the ``energy of events'',
and postulating the vacuum state with the minimal energy. One can
consider how this WDW SR/GR correspondence with the vacuum
postulate solves the problems of cosmological singularity, initial
data and arrow of the time in the Hamiltonian cosmology.

The vacuum postulate restricts the motion of the Universe in the
field space of events and it
 means that for positive energy of events
 $P_\vh\geq 0$ the Universe moves forward $\vh>\vh_I$,
 and for negative $P_\vh\leq 0$,
 moves backward $\vh<\vh_I$,
 where $\vh_I$ is the initial data. In quantum theory $\vh_I$ is treated
 as a point of creation of the Universe with positive energy $P_\vh\geq 0$,
  or
as a point of annihilation  of the anti-Universe with positive
energy (when the energy of events decreases $P_\vh\leq 0$). We can
see that the point of singularity $\vh=0$ belongs to the
anti-Universe: $P_\vh<0$. The Universe with the positive energy of
events does not contain the cosmological singularity $\vh =0$.

 This WDW analogy of the Universe with a relativistic
 particle allows us to get the causal Green function \begin{multline}\label{g}
 G(\vh_I|\vh)=G_{+}(\vh_I| \vh)\Theta(\vh-\vh_I)+\\+G_{-}(\vh_I|
 \vh)\Theta(\vh_I-\vh), \end{multline} where $G_{-}(\vh_I| \vh)=G_{+}(\vh|
 \vh_I)$
 is the probability to find the Universe at the point
 $\vh$, if the Universe was at the point $\vh_I$.

 If we quantize the constrained system (\ref{ham}) after
 the solution of the constraint $P_\vh=\pm E$, the Green function
 $G_{+}(\vh_I|\vh)$ satisfies the linear version of the WDW equation \be
 (\hat{P}_\vh-E(\vh))G_{+}(\vh_I|\vh)=0;\qquad \hat{P}_\vh=-i
 d/d\vh.
 \ee In this case a solution of this equation can be
 written in the form
 \be
 \label{fie}
 G_{+}(\vh_I|\vh)=
 \exp\left\{ i \int^\vh_{\vh_I}d\bar{\vh}E(\bar{\vh})\right\}.
 \ee

The status of conformal time $\eta$ in the Hamiltonian cosmology
follows from the
 variation of the action (\ref{ham}) with
 respect to the momentum $p_\vh$ that gives $p_\vh=2V_0\vh'$ ($\vh'=d\varphi/d\eta$).
 The substitution of this equation into
   the energy constraint  (\ref{conn}) leads to a
 conformal version of the  Freedman equation 
 \be\label{cfe} \vh{'}{}^2=\rho_0(\vh);\qquad
 \vh{'}=\frac{d\vh}{d\eta};~~~d\eta=N_0dx^0, \ee
 where $\eta$ is the
 conformal time. The solution of (\ref{cfe})
 \be\label{et}
 \eta(\vh_{I}, \vh)=\pm\int^\vh_{\vh_I}\frac{d\vh}{\sqrt{\rho_0(\vh)}}
 \ee
 admits any sign and values of $\eta$ and $\vh>0$, besides the
 point singularity $\vh=0$.

 It is easy to show that the vacuum postulate
 leads to the arrow of the conformal time (\ref{et})
 $\eta(\varphi_I,\varphi)>0$:
for both the Universe $P_\vh>0,
 \vh>\vh_I$ and the anti-Universe $ P_\vh<0, \vh_I<\vh $.

Thus, the treatment of the cosmological scale factor
 as the time-like
 dynamic variable (and  its canonical momentum
 as the evolution generator of motion in the space of events
 restricted by the vacuum postulate)
  gives us a possibility to solve the
 topical problems of cosmological singularity, the Hubble evolution,
 arrow of time, and
 initial data (see Appendix A).

\section{\label{chr}HAMILTONIAN GENERAL RELATIVITY WITH COSMOLOGICAL DYNAMICS}

\subsection{Separation of cosmological scale factor in GR}

 SR and cosmology gave us the set of arguments in the favor of
 the consideration of the evolution parameter as a cosmological
 scale factor in GR.
 The cosmological scale factor can be  included in the theory (\ref{t1})
 by the conformal transformations of  all fields with the conformal
 weight $(n)$: $F^{(n)}=a^n\bar{F}^{(n)}$, including the Cartan forms
 \be\label{csf}
 \omega_{(\alpha)}=a\bar{\omega}_{(\alpha)},~~~~\psi^2=a\bar \psi^2.
 \ee
 It is just the definition of the cosmological perturbation theory
 \cite{bard,kodam}, if we substitute this conformal transformation into
 equations of motion. It is logically correct, if
 the scale factor is treated as an external field parameter.

 However, in our case of the Hamiltonian approach to GR
  the scale factor will be considered as a dynamic variable, to convert it
 into the {\it dynamic evolution parameter}.

 There is an essential
 difference between an external field parameter and an
 internal dynamic variable. In the first case we can consider the
 theory on the level of equations of motions. In the second case,
  to determine the complete set of canonical momenta, the scale factor
  should be introduced into the GR action as a dynamic variable.
 The substitution of the transformation (\ref{csf}) into the action (\ref{t1})
 leads to the expression
 \be \label{sv1}S[\varphi_0|F]={S}[\varphi|\bar{F}]+ V_0\int
 dx^0 \varphi(x^0)\partial_0\left(\frac{\partial_0 \varphi}{N_0}\right),
 \ee
 where ${S}[\varphi|\bar{F}]$ is the sum of the initial GR action (\ref{l})
 \be\label{l11} S_{GR}[\vh|e]=\int dx^0d^3 x \left({\bf K}[\vh|e]-{\bf P}[\vh|e]
 +{\bf S}[\vh|e]\right)
 \ee
 and the SM one (\ref{t1}) with
  the running masses including the
  Planck mass $\varphi=\varphi_0 a$, $V_0=\int d^3 x$ is
 the volume of the Dirac coordinate space,
 \be \label{n0}
 {N_0(x^0)}^{-1}={V^{-1}_{0}}\int_{V_0}d^3x{\bar N^{-1}_d(x^0, x^i)}
 \equiv\left\langle{\bar N^{-1}_d}\right\rangle
 \ee
 is the averaging of the corresponding inverse Dirac lapse function over the
 spatial volume. The averaging lapse function $N_0(x^0)$ determines the
 geometric  time $\zeta$
 \be\label{ght}
d\zeta=N_0(x^0)dx^0.
 \ee

After the substitution of (\ref{csf}) into the action its spatial
determinant part (SDP) takes the form  (to surface terms)
 \begin{multline}\label{sdp}
 S_{\rm SDP}=-V_0\int dx^0
 \left[{\vphantom{\int}}4\vh^2\langle {\bar N_d\bar \pi}^2_{\psi}\rangle
 +\right.\\ \left.+4\vh\partial_0\vh\langle { \bar \pi}_{\psi}\rangle+
 (\partial_0\vh)^2\langle { \bar N_d}^{-1}\rangle{\vphantom{\int}}\right],
 \end{multline}
 where the first term arises from the kinetic part ${\bf K}[\vh|e]$,
 the second goes from the "surface" one ${\bf S}[\vh|e]$, as it is
 not the total derivative if the constant $\vh_0$ is replaced by
 the scale factor after the conformal transformation, and the
 third term is the  action for the scale factor,
 \be\label{proi1}
 \bar \pi_{\psi}={\bar N^{-1}_d}\left[
 (\partial_0-N^l\partial_l)\log{\bar
 \psi}-\frac16\partial_lN^l\right]
 \ee
 is the velocity deviation of logarithm of the spatial determinant
  \be\label{dev}
  \log \psi^2=\log a(x^0)+\log {\bar \psi}^2.
  \ee
 The scale factor is a dynamic variable, its  canonical
 momentum can be obtained by the variation of the Lagrangian  (\ref{sdp})
 with respect to  velocity $\partial_0\varphi$
\be\label{P}
 P_\vh\equiv
  \frac{\partial {L}_{SDP}}{\partial(\partial_0 \vh)}=
 -2V_0\vh'-4V_0\vh\langle { \bar \pi}_{\psi}\rangle,
 \ee
 while the averaging of the canonical momentum of the spatial
 determinant is
 \be\label{pi}
  \langle { \bar p}_{\psi}\rangle\equiv \langle
  \frac{\partial {\cal L}_{SDP}}{\partial(\partial_0 \log \bar \psi)}\rangle=
 -8\vh^2\langle { \bar \pi}_{\psi}\rangle-4\vh\vh',
 \ee
 here after $\varphi'={d\varphi}/{d\zeta}$.
 It is easy to convince that the canonical momenta
  $p_\alpha=[P_\vh,\langle { \bar p}_{\psi}\rangle]$ could not
  be expressed in terms of the velocities
  $\pi_\alpha=[\vh', \langle { \bar \pi}_{\psi}\rangle]$ as
  the corresponding set of equations
  \be\label{de1}
  p_\alpha=D_{\alpha\beta}\pi_\beta,
  \ee
  where the matrix
  \be D_{\alpha\beta}=
  \begin{pmatrix}
 D_{11},& D_{12},\\
 D_{21},& D_{22},
  \end{pmatrix}=
  \begin{pmatrix}
 -2 V_0,& -4\vh,\\
 -4 V_0\vh,& -8\vh^2,
 \end{pmatrix}
 \ee
 has the zero  determinant
 $|D_{\alpha\beta}|=16V_0\vh^2-16V_0\vh^2=0$.

 This means that the action (\ref{sdp}) is singular due to
 the double counting of the spatial determinant variable.

 To remove the double counting, the field variable $\log \bar \psi$ in
  Eq. (\ref{dev}) should be
  defined in the class of functions  distinguished  by the strong constraints
 \be\label{h1}
 \int_{V_0}d^3 x \log \bar \psi\equiv 0,~~~~\int_{V_0}d^3 x\bar{\pi}_\psi\equiv 0.
 \ee
 These constraints are nothing but the orthogonality of the scale
 factor and its velocity to the deviation of the spatial determinant
 logarithm $(\log\bar \psi)$.

 After that the action (\ref{sv1}) takes the form
 \begin{multline} \label{sv11}
 S[\varphi_0|F]= \int dx^0\left\{
 V_0\varphi(x^0)\partial_0\left(\frac{\partial_0 \varphi}{N_0}\right)
 +\right.\\ \left.+\int d^3x \left({\bf K}[\vh|\bar e]-{\bf P}[\vh|\bar e]+
 {\cal L}_{\rm SM}\right)\right\},
 \end{multline}
 where ${\cal L}_{\rm SM}$ is the Lagrangian density of the SM model,
$N_0$ is defined by eq. (\ref{n0}), and ${\bf K}[\vh|{\bar e}]$
and ${\bf P}[\vh|\bar e]$ are
 given by eqs. (\ref{k}) and (\ref{p}) respectively, where $\vh_0,e$
 are replaced by $\vh,\bar e$.

 In this case, the scale factor momentum   $P_\vh$ is
 completely separated from the local momentum $\bar{p}_\psi$,
 which
 satisfies the week Dirac constraint of the minimal surface:
 \be\label{h3}
 \bar{p}_\psi \simeq 0.
 \ee
 This separation allows us to get a
 version of the Friedmann equations in the exact theory.

\subsection{\label{em}Friedmann-like equations in exact theory}

 The equation of the lapse function
  \be\label{ladm00g}
 \bar N_d\frac{\delta S[\varphi_0|F]}{\delta \bar N_d }=0\ee
takes the form
\be \frac{\vh'^2}{\cal N}={\cal N}{\cal H}_t,
 \ee
 where
 \be\label{n}
 {\cal N}=\frac{\bar N_d}{N_0}
 \ee
 is the reparametrization invariant part of the lapse
 function  (\ref{n0}) satisfying the constraint
 $\langle { \cal N}^{-1}\rangle=1$  and
 \be\label{ht}
 {\cal H}_t={\cal H}_d(\vh|\bar e) +{\cal H}_s
 \ee
 is the sum of the Hamiltonian density (\ref{hd}),
 where $\varphi_0, e$ are replaced by $\varphi,\bar e$:
 \be\label{hdn}
 {\cal H}_d(\vh|\bar e)= \frac{6p_{(ab)}p_{(ab)}}{\vh^2}
 +
 \frac{\vh^2\bar \psi^{8}}{6} \left[{}^{(3)}R({\bf e})+
 \frac{8\Delta\bar \psi}{\bar \psi}\right],
 \ee
  and \be\label{hs}
 {\cal H}_s=\bar \psi^{12}T^0_{0({\rm SM})}.
 \ee
 is Hamiltinian density  of the Standard Model fields given by the
  zero-zero component of the energy momentum tensor
  $T^\mu_{\nu({ \rm  SM})}$

 Averaging  Eq. (\ref{ladm00g}) over the volume $V_0$ leads to
  the Friedmann-like equation in the exact theory
 \bea\label{f0}
 \langle\bar N_d\frac{\delta S[\varphi_0|F]}{\delta \bar N_d }\rangle=0&
 |\!\!\models\!\!\!\!\!\!
 \Longrightarrow&\vh'^2=\rho_t,
 \eea
 where
 \bea\label{1f0}
 \rho_t\equiv \langle {\cal N}{\cal H}_t\rangle
 \eea
 is the total generator of evolution under the geometric time
 $\zeta$  (\ref{ght}) of all dynamic variables except  the scale factor.

 The second equation of the Friedmann cosmology
 is obtained by the variation of the action (\ref{sv11})
 with respect to the scale factor $\vh$
 \bea\label{fk}
 \vh\frac{\delta S[\varphi_0|F]}{\delta \vh }=0&
 |\!\!\models\!\!\!\!\!\!
 \Longrightarrow&2\vh\vh''=\rho_t-3p_t,
 \eea
 where
 \be
 3p_t\equiv
 \langle { 3{\bf K}[\vh|\bar e]-{\bf P}[\vh|\bar e]+2{\bf S}[\vh|\bar
 e]+{\cal N}\psi^{12}T^k_{k({  \rm  SM})}
 \rangle}
 \ee
 is the exact pressure of all fields including the SM ones.
 Equations (\ref{f0}) and (\ref{fk}) give the relation
 \bea\label{1fk}
 2\vh'^2 + 2\vh\vh''\equiv (\vh^2)''=3(\rho_t-p_t)=
 \langle{{\hat {\bf A}}_t{\cal N}}\rangle,
 \eea
 where the expression
 \begin{multline}\label{3fk}
 {\hat {\bf A}}_t{\cal N}\equiv 4{\bf P}[\vh|\bar e]-2{\bf S}[\vh|\bar
 e]+\\ +\psi^{12}\left(3T^0_{0({  \rm  SM})}
 -T^k_{k({  \rm  SM})}\right){\cal N}
 \end{multline}
 determines
 the equation of $\log\bar \psi$ (\ref{d01}) added by the SM
 fields:
 \bea\label{4f2}
 \frac{1}{2}\frac{\delta S[\varphi_0|F]}{\delta \log \bar \psi }=0&
 |\!\!\models\!\!\!\!\!\!
 \Longrightarrow&{{\hat {\bf A}}_t{\cal N}}=\langle{{\hat  {\bf A}}_t{\cal N}}\rangle.
 \eea

 The second equation for the deviations from the average can be obtained by the
 substitution of (\ref{f0}) into  (\ref{ladm00g})
 \bea\label{5f0}
 \frac{\langle {\cal N}{\cal H}_t\rangle}{\cal N}
 &=&{\cal N}{\cal H}_t.
\eea
  In the infinite volume limit
 $V_0\to\infty$, Eqs. (\ref{5f0}) and (\ref{4f2}) are converted into
 the zero Hamiltonian density ${\cal H}_t=0$ and the Gribov
 zero ${\hat {\bf A}}_t{\cal N}=0$ because the averages
  $\langle  {\cal H}\rangle $ and  $\langle  {\hat {\bf A}}_t{\cal N}\rangle $
  are equal to zero.
 In the case of a finite volume the paradoxes of the coordinate time
 evolution considered in Section 2 can be removed by the
  change the order of
 the infinite volume limit and variation of the action like in QFT and statistical
 physics.

\subsection{\label{hr1}Cosmological geometro - dynamics}

 All equations considered above can be reproduced by varying the action in
   the Hamiltonian approach
 \begin{multline}\label{12ha1}
 S[\varphi_0|F]=\int dx^0\left\{-P_{\vh}\partial_0\vh+
 N_0\frac{P^2_\vh}{4V_0}+\right.\\ \left.+\int d^3x
 \left[\sum\limits_{  F}P_{  F}\partial_0F
 +{\cal C}-N_d{\cal H}_t\right]\right\},
 \end{multline}
 where $P_{  F}$ is the set of the field momenta, $N_d=N_0{\cal
 N}$, and
 \be\label{2ha3}
 {\cal C}=N_{(a)} {\cal P}_{t(a)} +C_0p_\psi+ C_{(a)}\partial_k{\bf e}^k_{(a)}
 \ee
is the sum of constraints. In this Hamiltonian approach
 the expressions ${\cal H}_t$ (\ref{ht}) and ${\hat A}_t$ (\ref{3fk})
 do not depend on ${\cal N}$.

 Recall that in the standard Hamiltonian approach without the scale
 factor considered in Section \ref{hr}
 we have the energy constraint ${\cal H}_{\rm t}=0$, whereas in
 the
 scheme of
 the Hamiltonian approach with the scale factor  we get
 two energy constraints (as we have seen above in Eqs. (\ref{f0})
 and (\ref{5f0})):  the global
 \be\label{ha5}
 \frac{P^2_\vh}{4V_0}=  H_{\rm t}\equiv\int d^3x{\cal N}{\cal H}_t
 \ee
and the local one  (\ref{5f0}).
 The  global constraint (\ref{ha5}) defines the effective Hamiltonian
\be\label{ha7}
 P_\vh= \pm 2 \sqrt{V_0H_{\rm t}}.
 \ee
 treated as a generator of
 evolution of all physical fields $F$
 in the field space of events $[\vh|F]$ with respect to the field
 evolution parameter $\vh$.
 The local constraint  (\ref{5f0}) means that only  nonzero
harmonics of the local energy density are equal to zero in
perturbation theory.

Solutions of the equations of the theory (\ref{12ha1}) in terms of
the geometric time (\ref{ght}) determine the geometric interval
(\ref{ds}), (\ref{adm}), (\ref{psi}) \be
ds^2=\omega^2_{(0)}-\omega^2_{(a)}, \ee where \bea
\omega_{(0)}&=&a(\zeta){\psi}^6 {\cal N}d\zeta, \\
\omega_{(a)}&=&a(\zeta)\psi^2({\bf e}_{(a)i}dx^i+N_{(0)}d\zeta).
\eea Recall that the invariant geometric time $\zeta$  (\ref{ght})
is defined by the Friedmann-like equation in the exact theory
 (\ref{f0}):
\be \label{rt}\frac{d\varphi}{d\zeta}\equiv\varphi'=
\pm\sqrt{\rho_t(\varphi)}. \ee The solution of (\ref{rt}) \be
 \label{111}\zeta(\varphi_0|\varphi_I)=\pm\int_{\vh_I}^{\vh_0}
 \frac{d\vh}{\sqrt{\rho_t(\varphi)}}
 \ee
can be considered as a pure relativistic relation between the
evolution parameter $\vh$ and geometric time  (\ref{ght}).
Equations (\ref{fk}) and (\ref{rt}) describe Friedmann-like
cosmology without any assumption about homogeneity as  pure
relativistic effects of the Hamiltonian description of GR in the
field space of events.

\subsection{\label{hr2}Hamiltonian reduction and the red shift representation}

 In this approach Eq. (\ref{5f0})  can be solved
 immediately:
\bea\label{6f0}
 {\cal N}&=&\frac{{\langle \sqrt{\cal H}_t\rangle}}{\sqrt{{\cal H}_t}}.
\eea This solution corresponds to the positive energy of events
(\ref{ha7})
 \be\label{1P}
 P_{\vh(+)}=2V_0\vh'=2V_0{\langle \sqrt{\cal H}_t\rangle}.
 \ee
 The substitution of Eq.  (\ref{6f0}) into Eq.  (\ref{12ha1}) leads
 to the reduced Hamiltonian action
 \begin{multline}\label{2ha2} S_{(+)}[\varphi_I|\varphi_0]|_{\rm   energy
 ~constraint}=\\
 =
 \int\limits_{\vh_I}^{\vh_0} d\vh\left\{\int d^3x
 \left[\sum\limits_{  F}P_{  F}\partial_\vh F
 +\bar{\cal C}-2\sqrt{{\cal H}_t}\right]\right\}
 \end{multline}
 like Eq. (\ref{srr}) in SR, here $\vh_I$ is a point of the
 Universe creation,
 $\bar{\cal C}={\cal
 C}/\partial_0\vh$ and the scale factor $\vh$ plays the role
 of a dynamic evolution parameter in the space of events $[\varphi|F]$.
  One can be convinced that
 varying this reduced action with respect to $\log\bar \psi$
 copies  Eq. (\ref{4f2}) where $\cal N$
  is determined by Eq. (\ref{6f0}). The action (\ref{2ha2}) gives
  the evolution of fields directly in terms of the red shift parameter
  connected with
   the scale factor $\vh$ by the relation $\vh=\vh_0/(1+z)$.

 The local energy density ${\cal H}_t$  (\ref{ht}) can be given as
 a sum
 of the homogeneous cosmological density (considered in the action
 (\ref{ham}))
  and the local density of a particle-like excitations
 \be \label{321}{\cal H}_t=\rho_0(\vh)+\eu{H}_{part}.
 \ee
 Using the nonrelativistic decomposition of the square root
  $\sqrt{{\cal H}_t}$ in
 the reduced action (\ref{2ha2})
 \begin{multline}\label{2h3} 
 \int\limits_{\vh_I}^{\vh_0} d\vh\int d^3x
 2\sqrt{\rho_0(\vh)+\eu{H}_{part}}=\\
 =
 \int\limits_{\vh_I}^{\vh_0}d\vh \int d^3x
 \left[2\sqrt{\rho_0(\vh)}+
 \frac{\eu{H}_{part}}{\sqrt{\rho_0(\vh)}}\right]+...
 \end{multline}
 and the definition of the conformal time (\ref{et})
  $d\eta=d\vh/\sqrt{\rho_0(\vh)}$( that coincides in the approximation
  with the geometric one $\zeta$) one can obtain the reduced action
  (\ref{2ha2}) in the form of the sum
 \be\label{2h4}
 S_{(+)}[\varphi_I|\varphi_0]|_{\rm   energy
 ~constraint}=S_{\rm c}+S_{\rm part}+\ldots, \ee
 where the first term is  the reduced cosmological action (\ref{ham})
 and the second is an ordinary action of particle excitations in terms
 of the conformal time
\be\label{2h5} S_{\rm part}=
 \int\limits_{\eta_I}^{\eta_0} d\eta\int d^3x
 \left[\sum\limits_{  F}P_{  F}\partial_\eta F
 +\bar{{\cal C}}-{\cal H}_{part}\right]
 \ee
 with the running masses $m(\eta)=a(\eta)m_0$,
 that describe the cosmological creation of particles \cite{ps1}.
 {Note that in quantum field
 theory the interaction is separated at first
$\sqrt{\rho_0+{\cal H}}$=$\sqrt{\rho_0+{\cal H}_{(2)}}+
 {\cal H}_I/\sqrt{\rho_0+{\cal H}_{(2)}}$, which leads to a form factor
 decreasing the ultraviolet divergences \cite{pp}.}

 \subsection{\label{cpt}Hamiltonian cosmological perturbation theory}

Let us compare on the classical level the Hamiltonian cosmological
perturbation theory with the conventional one \cite{lif,bard},
where the cosmological factor is considered as an external field
with double counting of the spatial determinant. The cosmological
perturbations of the metric components
 \begin{gather}
 {\cal N}=(1- \overline{\Phi}_N-3\overline{\Phi}_h),~~~ \overline{\psi}=
 \left(1+\frac{\overline{\Phi}_h }{2}\right),\\
 d{\bf e}_{(a)i}=dh^{(TT)}_{(a)i},
  \end{gather}
  \be\label{shift}
 N_{(a)}=\partial_{(a)}\sigma+N^{T}_{(a)};~~~~\partial_{(a)}N^{T}_{(a)}=0,
 \ee in the new
cosmological perturbation theory
\bea &&\omega_{(0)}=a(\eta)(1- \overline{\Phi}_N)d\eta,\\
&&\omega_{(a)}=a(\eta)(1+\overline{\Phi}_h)\times \nonumber\\
&&\times(dx^i_{(a)}+h^{(TT)}_{(a)i}dx^i+
\partial_{(a)}\sigma
d\eta+N^{(T)}_{(a)}d\eta),\eea
 are defined in the class of functions with the nonzero Fourier harmonics
\be \widetilde{\Phi}(k)=\int d^3x \overline{\Phi}(x)e^{ikx} \ee
  satisfying the strong
 constraint
 $\int d^3x \overline{\Phi}(x)\equiv 0$.
In the same way one can decompose the energy-momentum tensor
components: \be T^0_{0}{}^{\phantom{rm}}_{\rm
sm}=\rho_s+\overline{\vphantom{\bigl|}T}^{(1)}_{00};~~~~~~~~
T^k_{k}{}^{\phantom{rm}}_{\rm
sm}=3p_s+\overline{\vphantom{\bigl|}T}^{(1)}_{kk}, \ee
 where $\rho_{  \rm s}$, $p_{  \rm s}$
 are the SM model density and pressure.
  The first
 order of the decomposition of expressions
  (\ref{k}), (\ref{p}), (\ref{s}) is
 \be\label{kps}
 {\bf K}^{(1)}=0,~~~{\bf P}^{(1)}=\frac{2\vh^2}{3}\triangle
 \overline{\Phi}_h
 ,~~~{\bf S}^{(1)}=-\frac{\vh^2}{3}\triangle\overline{\Phi}_N.
 \ee
In the approximation
$\overline{\Phi}_h,\overline{\Phi}_N\ll\rho_s, p_s\,$
 Eqs. (\ref{4f2}) and (\ref{5f0}) for the scalar components take
the form \begin{align}\label{107}
\widetilde{T}^{(1)}_{00}&=\frac{2\varphi^2k^2}{3}\widetilde{\Phi}_h+
2\rho_s\widetilde{\Phi}_N,\\
\widetilde{T}^{(1)}_{00}+\widetilde{T}^{(1)}_{kk}&=
-9(\rho_s-p_s)\widetilde{\Phi}_h+\notag\\\label{108}
 &+\left(\frac{2\varphi^2k^2}{3}-
5\rho_s+3p_s\right)\widetilde{\Phi}_N \end{align} added by the
Dirac minimal surface constraint
 \be
 \label{dg}
 \triangle\sigma=\frac{3}{4}\overline{\Phi}_h'.
 \ee
 In the Newton case: $p_s,
 \rho_s\ll{\varphi^2k^2}$,
 we obtain the standard classical solutions:
 \be\label{dgg}
 \widetilde{\Phi}_h=\frac{3}{2\varphi^2k^2}\widetilde{T}^{(1)}_{00};~~~~
 \widetilde{\Phi}_N
 =\frac{3}{2\varphi^2k^2}\left[\widetilde{T}^{(1)}_{00}
 +\widetilde{T}^{(1)}_{kk}\right].
 \ee
 For the tensor and vector components we got the equations
 \be\label{tiktt} \overline{T}^{TT}_{ik}=\frac{\varphi^2}{12}\left[-\triangle
 h^{(TT)}_{ik}+
 \frac{(\varphi^2{h^{TT}_{ik}}'){\vphantom{h^{TT}_{ik}}}'}{\varphi^2}\right];\ee
 \be(\partial_i T^{(TT)}_{ik}=T^{(TT)}_{ii}=0),\ee
\be\label{tok2}
  T^{0\:(T)}_k=-\frac{\varphi^2}{12}N^{T}_{(a)};~~~
 (\partial_kT_k^{0\:(T)}=0). \ee

 Eqs. (\ref{107}), (\ref{108}), (\ref{tiktt}), and  (\ref{tok2})
  determine six  components
  ($\overline{\Phi}_N$, $\overline{\Phi}_h$, $N^{T}_{(a)}$, $h^{TT}_{ik}$)
 of the metric in the Dirac gauge of the minimal surface
 (\ref{dg}) that determines the longitudinal component  $\partial_{(a)}\sigma$
 of the shift vector (\ref{shift}).

The Hamiltonian form of the cosmological perturbation theory does
not require its convergence to be proved because the perturbations
are in a different class of functions (with nonzero Fourier
harmonics) than the cosmological dynamics described by the exact
equations (\ref{1f0}), (\ref{fk}). In contrast to the standard
cosmological perturbation theory the Hamiltonian version contains
the shift of the coordinate
  origin in the process of evolution, and the Newton-like
  form of interactions appears after resolving
  the constraints.

\subsection{Cosmological generalization of the Schwarzschild solution}

The substitution \be {\cal N}_{\psi}=\psi^7{\cal N} \ee allows us
to extract the nabla operator
$\triangle=\partial_{(a)}\partial_{(a)}$ in Eqs. (\ref{4f2}) and
(\ref{5f0}):
$${{\hat{\bf  A}}_t{\cal N}}=\langle{{\hat{\bf A}}_t{\cal
 N}}\rangle;~~~~~ {\cal N}{\cal H}_t=\frac{\langle {\cal
N}{\cal H}_t\rangle}{\cal N},
$$ where
\bea {{\hat {\bf  A}}_t{\cal N}}&\equiv&4{\bf P}-{\bf
S}+\psi^5{\cal N}_{\psi}(3T^0_0-T^k_k)\\ {\cal N}{\cal
H}_t&\equiv& {\bf P}+{\bf K}+\psi^5{\cal N}_{\psi}T_0^0, \eea so
that the expressions for ${\bf P}$ and ${\bf S}$ take the form
\bea {\bf P}&=&\frac{4\varphi^2}{3} {\cal
N}_{\psi}\triangle\psi+\frac{\varphi^2}{6}{\cal
N}_{\psi}R(\textbf{e}),\\
{\bf S}&=&\frac{\varphi^2}{3}({\cal N}_{\psi}\triangle\psi-
\psi\triangle{\cal N}_{\psi}). \eea
 In the case when ${\bf P}$ and
${\bf S}$ are equal to zero, we come to the equations \be
\triangle{\cal N}_{\psi}=0,~~~~~ \triangle\psi=0,~~~~~ R({\bf
e})=0. \ee
 Solutions  of these equations are
 \bea \psi&=&1+\frac{r_{g}(\zeta)}{4r},\\
 {\cal N}_{\psi}&=&1-\frac{r_g(\zeta)}{4r},\eea
 where $r_{g}(\zeta)=M3/4\pi\vh^2$ is the gravitational radius.

 It is easy to see that the standard Schwarzschild metric
 in the vacuum $T^\mu_\nu=0$ can be treated as a solution
 of the Einstein equation in the approximation, where we
 neglect the dependence of masses on the geometric time:
 $r_g(\zeta)\sim r_g(\zeta_0)={\rm constant.}$
  The generalization
 of the standard Schwarzschild solution in conformal flat metric
 can be written in
 terms of the Cartan forms (\ref{adm}), (\ref{psi})
 \bea \omega_{(0)}&=&\frac{{\cal N}_{\psi}}{\psi}d\zeta, \\
 \omega_{(r)}&=&\psi^2\left(dr+\frac{V'}{r^2\psi^6}d\zeta\right),\\
 \omega_{(\theta)}&=&\psi^2r^2d\theta, \\
 \omega_{(\varphi)}&=&\psi^2r^2\sin\theta d\varphi; \eea here \be
 V(r,\zeta)=\int\limits_{}^{r} d\bar r~\bar r^2\psi^6(\bar r,\zeta) \ee
 determines the radial component of the shift vector satisfying
 the minimal surface constraint (\ref{h3}).

 \subsection{The vacuum postulate and the Faddeev -- Popov integral}
 In the considered version of GR with the vacuum postulate,
   the probability to find the Universe at the point
$(\vh_IF_I)$, if the Universe was created at the point $(\vh_0
F_0)$, is determined by the causal Green function similar to
expression (\ref{g})
  \begin{multline}\label{gg}
 G(\vh_IF_I|\vh_0 F_0)=G_{+}(\vh_IF_I| \vh_0
 F_0)\Theta(\vh_0-\vh_I)+\\+
 G_{+}(\vh_0F_0|\vh_IF_I)\Theta(\vh_I-\vh_0), \end{multline}
where
 \begin{multline} G_{+}(\varphi_IF_I|\varphi_0
 F_0)=\\=\int\prod_{x} \left[d\psi dN_{(a)}dC_{(a)}\prod_{F}
 \left(\frac{dP_{F}dF}{2\pi}\right)\right]\times\\ \times D \exp\{iS_{+}[\varphi_I|\varphi_0]\}, \end{multline}
 here $S_{+}$ is the reduced action given by Eq. (\ref{2ha2})),
 $C_{(a)}, N_{(a)}$ are the Lagrange factors (see Eqs.
 (\ref{12ha1}), (\ref{2ha3})), and $D$ is the Faddeev --
 Popov determinant: \be D= \det\hat{\bf A}_t det\hat{\bf
 B},\ee  $\hat{\bf A}_t$ and $\hat{\bf B}$ are defined by
 (\ref{3fk}) and (\ref{d03}).
We can see that the functional integral does not contain Gribov
ambiguity ($\hat{\bf A}_t{\cal N}\neq 0$) and zero-energy (${\cal
H}\neq 0$), and it is defined in terms of the invariant evolution
parameter $\varphi$. This functional integral and postulate about
the existence of physical vacuum as a state with the lowest energy
$(E=P_{\vh}>0)$ solve on the level of exact theory the topical
problems of cosmology: initial data $(F(\varphi=\varphi_I)=F_I)$,
arrow of time $(\eta>0)$, and cosmological singularity
$(\varphi\neq 0$, $\varphi>\varphi_I).$

This functional integral does not contradict  the Hamiltonian
cosmology (\ref{ham})
 (where the conformal time
 is an invariant under reparametrizations of the coordinate time
 and the scale factor is the internal dynamic variable), and can be
 considered as the generation functional of the Hamiltonian cosmological
 perturbation theory presented in Subsection \ref{cpt}.

\subsection{Discussion}
The Hamiltonian dynamics of GR was formulated
 \cite{dir,ADM,berg,shw,fadpop} by analogy with the Newton theory of
 nonrelativistic particle considered as a representation of the Galilei group.
In the
 present paper, we try to formulate the Hamiltonian  GR theory
 by analogy with the theory of a relativistic particle
 formulated as a construction of unitary irreducible representations
 of the Poincare group. We can find all elements of this construction
 in the Hamiltonian cosmological perturbation theory:
 the field space of events $[\varphi|F]$ containing,
 time-like variable $\varphi$,
 its canonical momentum $P_{\varphi}$ as the evolution Hamiltonian,
 the
  vacuum postulate, and
the separation of observables into the dynamic sector and the
geometric one.

  In contrast to the Newton mechanics
  the theory of a relativistic particle (SR)
  contains the coordinate  nonmeasurable evolution parameter and
  two measurable evolution parameters: the time as a dynamic
  variable and  the time as a geometric interval (see Fig.1).

\begin{figure}[!hbt]
  \begin{center}
  \includegraphics[width=8cm,height=5cm]{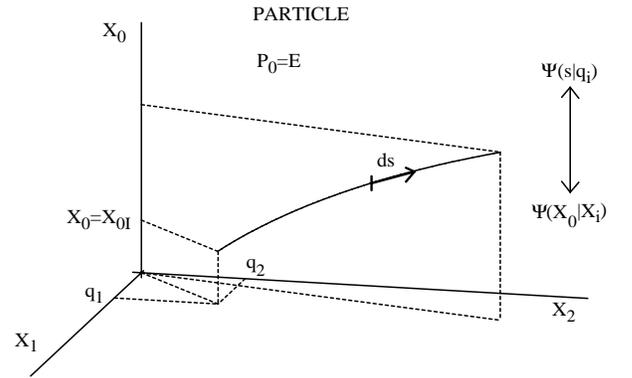}\nopagebreak
  \end{center}
  \caption{The world line $[s]$ of a relativistic particle in the
  space of events [$X_0| X_i$] with the initial data [$X_{0I}| q_I$]
  treated as the point of creation of the particle.}
  \vspace{0.5cm} \label{part}
 \end{figure}

\begin{figure}[!hbt]
  \begin{center}
  \includegraphics[width=8cm,height=5cm]{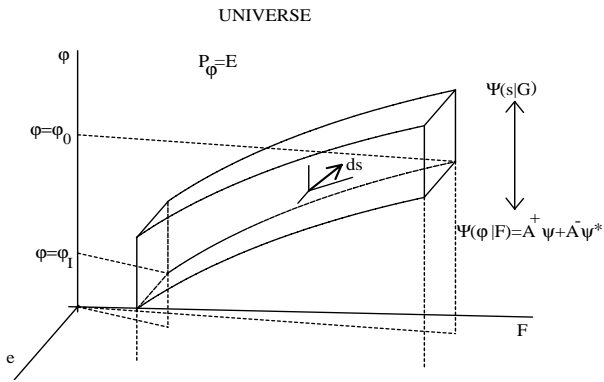}\nopagebreak
  \end{center}
  \caption{The world hypersurface [$s$] of the Universe in the space of
  events [$\varphi|F,e_{a\mu}$] with the vacuum initial data
  [$\varphi_I|G,e^I_{a\mu}$] treated as the point of creation of the Universe.}
  \vspace{0.5cm} \label{uni}
 \end{figure}

 The WDW SR/GR correspondence allows one to consider the Universe
 as an ordinary
 physical object given in the space of events in
 specific frame of reference similar to a relativistic particle
 given in the Minkowski space (see Figs. \ref{part}, \ref{uni}).

 The SR/GR correspondence means that we should
 point out the time-like dynamic variables in a specific  Lorentz
 frame and separate all measurable quantities
into the dynamic sector and the geometric one.

  The ``equivalent'' unconstrained
 Hamiltonian theory obtained by resolving a constraint
 can describe only the dynamic sector.

 Expression (\ref{gg}) determining the probability of the
 creation of the world hypersurface in the space of events
 together with the geometric interval in a specific frame of reference
 solves the problems of the Dirac formulation \cite{dir,fadpop}:
 the nonlocalizable energy, arrow
 of time, comological singularity, and the initial data.

 Faddeev and Popov proved in \cite{fadpop, popkon} that
 the integral  (\ref{z}) is not equivalent to the one in the relativistic
 invariant harmonic gauge $\partial_{\mu}(\sqrt{-g}g^{\mu\nu})=0$,
 which does not depend on a frame of reference. This means that
 the  functional integral  with the minimal surface (\ref{gg})
 depends on the frame of reference, and the problem of relativistic
 invariance of the Hamiltonian formulation arises.

 It is worthwhile to recall that the same problem arises also in SR,
 where
  in accordance with the theory of unitary irreducible representations
  of the Lorentz and Poincare groups
  the relativistic invariance means
 the invariance of a complete set of frames of reference with
 respect to the Lorentz transformations (see \cite{shwb}). In our case,
 the functional
 integral (\ref{gg}) should be repeated in each frame of reference of
 the complete set, so that  the complete set of
 functional integrals with minimal surfaces
 is relativistic invariant.

\section{\label{a} Conclusion}

 We investigated General Relativity under the supposition that
 the evolution parameter of its Hamiltonian description
 coincides with the cosmological scale
 factor.
 The resulting Hamiltonian  theory added by the vacuum postulate
 becomes free from the defects of the standard Hamiltonian
 approach and contains the cosmological Friedmann-like sector.
   The obtained Hamiltonian  cosmological theory
 differs from the standard Lifshits-Bardeen
 cosmological perturbation theory \cite{lif,bard,kodam},
 where the cosmological scale
 factor is treated as an external field with the double counting of
 the spatial metric determinant.
 In contrast to the standard cosmological perturbation theory
the Hamiltonian version contains the shift of the coordinate
  origin in the process of evolution, and the Newton-like
  form of interactions appears after resolving the constraints.

 It is interesting to apply this  cosmological Hamiltonian
 approach to GR  for the description
 of the CMB fluctuations.

\vspace{2cm}

 ACKNOWLEDGMENTS

 We are grateful to D. Blaschke, A. Gusev, P. Flin, P. Fomin,
 L. Lipatov and D. Mladenov for fruitful discussions.

\section*{ Appendix A: Field nature of time}

\renewcommand{\theequation}{A.\arabic{equation}}

\setcounter{equation}{0}

 To introduce the conformal time $\eta$  as a new field variable
 and its nonzero momentum
 as a proper energy of the geometric space of events,
  we can use the Levi-Civita-type
 canonical transformation \cite{bpp, lc}:
 $(P_\vh| \vh)\rightarrow(\Pi|\eta)$ to
 convert the energy constraint into a new canonical momentum $\Pi$.
 We consider this transformation using as an example the
 case of the Universe filled in by photons when $\rho_0(\vh)=\mbox{\rm const.}$
 In this case, this transformation takes the form
 \be
 P_{\varphi}=\pm2\sqrt{\Pi V_0},~~~~
 \varphi=\pm\frac12\sqrt{\frac{\Pi}{V_0}}\eta.
 \ee

 The action (\ref{ham}) becomes
 \be\label{1ham}
  S_{c}=\int dx^0\left[-\Pi\partial_0\eta+
  N_0\left(\Pi-\rho_0 V_0\right)\right].
 \ee
 After the reduction the non-zero energy
 $
 \Pi=V_0\rho_0
 $
 corresponding to the invariant
 conformal time  appears. The reduced action takes
 the form
 \be S=V_0\int\limits_{\eta_I}^{\eta_0}
 d\eta\rho_0=V_0\rho_0(\eta_I-\eta_0).
 \ee
 In quantum theory, where $\Pi=d/id\eta$,
  the geometric evolution is described by the
 wave function
 \be \label{geo}
 \psi_{\rm geometric}(\eta)=e^{iV_0\rho_0(\eta_I-\eta_0)}.
 \ee
 The Hubble evolution $\vh=\vh(\eta)$ is treated
 as a pure relativistic effect of
 the relation between two supplementary descriptions of the
 relativistic Universe by means of
 two wave functions: the field (\ref{fie}) and
 the geometric (\ref{geo}) ones.

\section*{ Appendix B: Central gravitational fields}

\renewcommand{\theequation}{B.\arabic{equation}}

\setcounter{equation}{0}

 Let us consider  the central gravitational
 field produced by a single mass object
 \be
 \overline{T}_{00}=M\left[\delta^3(x)-V_0^{-1}\right], ~~\overline{T}_{kk}=0.
 \ee
 (see Eqs. (\ref{dg}), (\ref{dgg})).
 Equation (\ref{dgg}) can be transformed into the integral form
 \be\label{fi}
  \overline{\Phi}(x)=\frac{3}{4\pi\varphi^2}\left.\int d^3
 x\frac{T_{00}}{|y-x|}\right|_{T_{00}=M\delta^3(x)}=\frac{r_g}{r},
 \ee
  where $r=\sqrt{x_1^2+x_2^2+x_3^2}$ and
  \be r_g=\frac{3M}{4\pi\varphi^2}=2GM;
  \ee
  here by
definition $\varphi=\varphi_0a,~~M=M_0a$ and
$\varphi_0=\sqrt{3/8\pi G}$.

  In the case of $T_{kk}=0$, it follows  from Eq. (\ref{dg}) that
  the shift vector $N_i$ is
 \be\label{ni}
 N^i=\left(\frac{3r_g'}{4}\right)\frac{x^i}{r}. \ee
 After substitution of the solutions (\ref{fi}) and (\ref{ni})
  into the conformal interval
  we have
  \be
  ds^2_c=\left(1-\frac{r_g}{r}\right)d\eta^2-
 \left(1+\frac{r_g}{r}\right)
 \left(dx_i+\frac32\frac{x^i}{r}r'_{g}d\eta\right)^2;
 \ee
 here $ds^2_c=ds^2/a^2(\eta)$.

 In the case of point mass distribution with the density \be\label{t00}
 T_{00}=\sum_I M_I\delta^3(z_I-x)
 \ee
the components of the metric $\overline{\Phi},~N^i$ are \be
\overline{\Phi}(x)=\sum^N_{I=1}\frac{r_{gI}}{|x-z_I|} \ee \be
N^i=\sum^N_{I=1}\frac{3r_{gI}'}{4}\frac{(x-z_I)^i}{|x-z_I|} \ee

  The conformal interval \be
 ds^2_c=(1-\overline{\Phi})d\eta^2-(1+\overline{\Phi})
 (dx^i+N^id\eta)^2 \ee
  determines an equation for the photon momenta
  \be
  p_{\mu}p_{\nu}g^{\mu\nu}\simeq (p_0+N^ip^i)^2(1+\overline{\Phi})
  -p^2_j(1-\overline{\Phi})=0,\ee
 from which we obtain
  \be
 p_0\simeq-N^ip^i+(1-\overline{\Phi})|p|;~~~~~~~ |p|=\sqrt{p^2_i}. \ee
 Finally, we obtain
 the relative magnitude of  spatial fluctuations of a photon energy
 in terms of  the metric components (the potential $\overline{\Phi}$ and
 shift function $N^i$) \be
 \frac{p_0-|p|}{|p|}=-(N^in^i+\overline{\Phi});~~~~~~~~~n^i=\frac{p_i}{|p|}.
  \ee
 The appearance of  spatial anisotropic
 fluctuations of the photon energy in the flow of photons is
 the consequence of the minimal surface  (\ref{ni}).


\begin{thebibliography}{}
\bibitem{dir}
 P. A. M. Dirac, Proc. Roy. Soc. {\bf A 246}, 333 (1958);
  Phys. Rev. {\bf 114}, 924 (1959).
 \bibitem{ADM}
 R. Arnowitt,  S. Deser, and  C .W. Misner, {Phys. Rev.} {\bf 116}, 1322 (1959);
{Phys. Rev.} {\bf 117}, 1595 (1960);
 {Phys. Rev.} {\bf 122}, 997 (1961).
 \bibitem{berg}
 P. Bergman, Rev. Mod. Phys. {\bf 33}, 510 (1961).
 \bibitem{shw}
 J. Schwinger, {Phys. Rev.} {\bf 130} 1253 (1963);
 {Phys. Rev.} {\bf 132}, 1317 (1963).
\bibitem{fadpop}
 L.D. Faddeev and V.N. Popov,  {Us.Fiz.Nauk} {\bf 111}, 427
 (1973).
 \bibitem{WDW}
 J. A. Wheeler,  {\it Lectures in
 Mathematics and Physics}
 (Benjamin,
 New York, 1968);
 B. C. DeWitt, Phys. Rev. {\bf 160}, 1113 (1967).
 \bibitem{M}
 C. Misner, {Phys. Rev.} {\bf 186}, 1319 (1969).
 \bibitem{Ryan1}
M. P. Jr. Ryan, L. C. Shapley,  ( {\it Homogeneous Relativistic
Cosmologies} (Princeton Series on Physics, Princeton University
Press, Princeton, 1975).
\bibitem{Ryan2}
M. P. Ryan,   {\it Hamiltonian Cosmology} (Lecture Notes in
Physics N 13, Springer Verlag, Berlin--Heidelberg--New York,
1972).
\bibitem{lif}
E.M. Lifshits, {Us.Fiz.Nauk}  {\bf 80}, 411 (1963), in Russian;
Adv. of Phys. {\bf 12}, 208 (1963).
\bibitem{bard}
J.M. Bardeen, Phys. Rev. {\bf D22}, 1882 (1980).
\bibitem{kodam}
H. Kodama, M. Sasaki, Prog. Theor. Phys., {N \bf 78}, 1 (1984).
\bibitem{fock29} V.A. Fock, Zs.f.Phys. {\bf 57}, 261 (1929).
\bibitem{og}
A.B. Borisov, V.I. Ogievetsky, Teor. Mat. Fiz., {\bf 21}, 329
(1974).
\bibitem{vlad}
A.L. Zel'manov, { Doklady AN SSSR} {\bf 227}, 78 (1976), in
Russian; Yu.S. Vladimirov, {\it Frame of references in theory of
gravitation} (M., Energoizdat, 1982), in Russian.
\bibitem{popkon}
N.P. Konoplyova, V.N. Popov, {\it Kalibrovochnye polya} (M.
Atomizdat, 1980), in Russian.

 \bibitem{shwb}
See the review of papers by V. Bargman, E.P. Wigner, and
A.S.Wightman in the monography by S. Schweber, Chapter 1, \S1, 4,
5. {\it An Introduction to Relavistic Quantum Field Theory} (Row,
Peterson and Co.  Evanston, III., Elmsford, N.Y. 1961).
\bibitem{18}
{The proper time $ds=e_{(1)}d\tau$ becomes
 dynamical variables in the geometric space of events $[s| q_i]$
 obtained by the Levi-Civita transformations
 $(p_{\mu}|x_{\mu})\Rightarrow (\Pi_{\mu}|s, q_{i})$, so that
 the constraint $p_{\mu} p^{\mu}-m^2=0$ becomes the new momentum
 $\Pi_0=0$ \cite{pp, bpp, lc, sh, gkp1, gkp}.}
\bibitem{pp}
M. Pawlowski, V.N. Pervushin, Int. J. Mod. Phys. {\bf 16}, 1715
(2001); [hep-th/0006116].
\bibitem{bpp}
B.M. Barbashov, V.N. Pervushin, M. Pawlowski, Phys. Particles and
Nuclei {\bf 32}, 546 (2001).
\bibitem{poi}
H. Poincare, C.R. Acad. Sci., Paris {\bf 140}, 1504 (1905).
\bibitem{ein}
 A. Einstein, Anal. d. Phys. {\bf 17}, 891 (1905).
\bibitem{ps1}
V.N. Pervushin, V.I. Smirichinski, J. Phys. A: Math. Gen. {\bf
32}, 6191 (1999).
\bibitem{lc}
T. Levi-Civita,  Prace Mat.-Fiz. {\bf 17}, 1 (1906).
\bibitem{sh}
S. Shanmugadhasan, {J. Math. Phys} {\bf 14}, 677 (1973).
\bibitem{gkp1}
S.A. Gogilidze, A.M. Khvedelidze and V.N.Pervushin, J. Math. Phys.
{\bf 37}, 1760 (1996); S.A. Gogilidze, A.M. Khvedelidze and V.N.
Pervushin, Phys. Rev. {\bf D 53}, 2160 (1996).
\bibitem{gkp}
S.A. Gogilidze, A.M. Khvedelidze and V.N. Pervushin, Phys.
Particles and Nuclei {\bf 30}, 66 (1999).
\bibitem{kasner} E. Kasner, Am. J. Math {\bf 43}, 217 (1921).
\bibitem{khal}
V.A. Belinsky, E. M. Lifshits, I.M. Khalatnikov, {Us. Fiz. Nauk}
{\bf 102}, 463 (1970), in Russian; JETF {\bf 60}, 1969 (1971), in
Russian; L.D. Landau, E.M. Lifshits, {\it The theoretical~
physics, {\bf V. 2.} The field theory} (M., Nauka, 1988).

\bibitem{114:a}
D. B. Blaschke, S. I. Vinitsky, A. A. Gusev, V .N. Pervushin, and
D. V. Proskurin, Physics of Atomic Nuclei {\bf 67},
 1050 (2004); [gr-qc/0103114].

\bibitem{PEPAN:34}
 B. M. Barbashov,  V .N. Pervushin, and
D. V . Proskurin, Phys. of Particles and  Nuclei, {\bf 34}, Suppl.
{\bf l},  S68 (2003).

\end{thebibliography}
\end{document}